\documentclass[]{interact}

\usepackage{epstopdf}
\usepackage{subfigure}

\usepackage[numbers,sort&compress,merge]{natbib}
\usepackage{graphicx}
\usepackage{braket}
\usepackage{gensymb}
\bibpunct[, ]{[}{]}{,}{n}{,}{,}

\theoremstyle{plain}

\theoremstyle{definition}

\theoremstyle{remark}

\begin{document}
\title{Attenuated Coupled Cluster: A Heuristic Polynomial Similarity Transformation 
       Incorporating Spin Symmetry Projection Into Traditional Coupled Cluster Theory}

\author{
\name{John A. Gomez\textsuperscript{a}\thanks{CONTACT John A. Gomez. Email: jag20@rice.edu} and 
   Thomas M. Henderson\textsuperscript{a,b} and
   Gustavo E. Scuseria\textsuperscript{a,b}}
   \affil{\textsuperscript{a}Department of Chemistry, Rice University, Houston, TX 77005, USA; \textsuperscript{b}Department of Physics and Astronomy, Rice University, Houston, TX 77005, USA}
}

\maketitle

\begin{abstract}
In electronic structure theory, restricted single-reference coupled cluster (CC)
captures weak correlation but fails catastrophically 
under strong correlation. 
Spin-projected unrestricted Hartree-Fock (SUHF), on the other hand, misses weak correlation but captures
a large portion of strong correlation. 
The theoretical description of many important processes, e.g. molecular dissociation,
requires a method capable of accurately capturing both weak- and strong correlation simultaneously,
and would likely benefit from a combined CC-SUHF approach.
Based on what we have recently learned about SUHF written as particle-hole excitations out of
a symmetry-adapted reference determinant, we here propose a heuristic coupled cluster doubles 
model to attenuate the dominant spin collective channel of the quadratic terms in the 
coupled cluster equations. Proof of principle results presented here are 
encouraging and point to several paths forward for improving 
the method further.
\end{abstract}

\section{Introduction}
We are interested in the simultaneous description of weak and strong correlation in electronic structure theory.  Weak, or dynamical, correlation is ubiquitous and is often conceptualized as electrons instantaneously avoiding one another.  The Hartree-Fock mean-field description of a weakly-correlated system is generally qualitatively correct, and a quantitative description can be provided by the coupled cluster family of methods.\cite{Paldus1999,Crawford2000,Bartlett2007,Shavitt2009} 
Strong, or static, correlation, on the other hand, is not as universally prevalent.  It typically arises from degeneracies in the system, and is associated with the qualitative failure of the
symmetry-adapted mean-field description.  
This failure, in turn, usually implies the breakdown of coupled cluster theory.

Strong correlation is often accompanied by the spontaneous breaking of a symmetry of
the Hamiltonian in the mean-field reference, e.g. the breaking of $S^2$ in the Hartree-Fock wavefunction
past the Coulson-Fischer point in a molecular dissociation curve. 
If one allows such symmetry breaking, e.g. unrestricted Hartree-Fock (UHF), then
broken-symmetry Hartree-Fock and coupled cluster
may give reasonable energies, but wrong wavefunctions, at least for finite systems
where symmetry breaking is artifactual, i.e. the result of approximations. 

Projected Hartree-Fock (PHF)\cite{Lowdin1955,RingandSchuck,Ripka,Schmid2004,Scuseria2011,Hoyos2012}
describes
strong correlation by restoring the symmetry-preserving component of 
the broken-symmetry mean-field reference. The PHF wavefunction is multideterminantal, but is compactly 
expressed as a linear combination of \textit{nonorthogonal} determinants. When 
expressed using \textit{orthogonal} determinants, the PHF wavefunction contains excitations to all orders, enabling
the PHF wavefunction to capture strong correlation. 
However, PHF generally misses a large amount of the weak correlation.

Since CC captures weak correlation but fails for strong
correlation, while PHF captures strong correlation but misses weak correlation, a natural question
is how to combine the two methods. One option is to perform CC atop a PHF wavefunction, and work
along these lines is presented elsewhere.\cite{Qiu2017}\cite{Jacob2017}
We here explore an alternative idea.
While PHF can refer to the projection of any symmetry of the Hamiltonian broken in the mean-field reference,
here, we are primarily concerned with the projection of $S^2$ out of an unrestricted Hartree-Fock
determinant (SUHF).
Although SUHF is traditionally written variationally, we have recently formulated SUHF for singlet states ($s=0$) as a 
polynomial similarity transformation (PoST) of particle-hole excitations out of a symmetry-adapted reference
determinant in the mathematical language of traditional coupled cluster.\cite{Qiu2016,Qiu2017}
In this work, we present attenuated coupled cluster (attCC), in which we use the PoST formulation of SUHF 
to inform a modificaton of the CCD amplitude equations that protects the method
from breakdown in the presence of strong correlation.
 
Modifying the coupled cluster equations and $T_2$ operator in order to describe strong correlation has a rich history, resulting in 
improved descriptions of ground states, excited states and properties.\cite{Paldus1984,Piecuch1990, 
Piecuch1996,Piecuch2001,Bartlett2006,Neese2009,Huntington2010,Small2012,Katz2013,Limacher2013,Stein2014,Henderson2014,Henderson2015,Bulik2015,Degroote2016}  
In previous work along these lines, we found that separating the singlet- and
triplet pairing channels of $T_2$ and isolating them from one another, giving singlet-paired (CC0)
and triplet-paired coupled cluster (CC1), protected CC from blowup.\cite{Bulik2015,Gomez2016a}
There, we decomposed $T_2$ along particle-particle/hole-hole (pp-hh), or ladder, channels,\cite{Scuseria2013}  eliminating 
the interaction of the channels completely. This result lead us to ask two questions: 1) can we protect
coupled cluster from breakdown by attenuating, rather than severing, the dialogue between offending
pairing channels? and 2) can we do so along particle-hole/particle-hole (ph-ph), or ring, channels?\cite{Scuseria2008} The present work is an attempt to answer these two questions.

In what follows, we first describe traditional coupled cluster theory and SUHF in both its standard variational and
our recently introduced PoST formulations.
We then introduce attenuated coupled cluster 
and present results on some small molecules and the Hubbard Hamiltonian.
Lastly, we remark on the combination of CCD with particle-number projection in the attractive pairing Hamiltonian before offering a concluding discussion.
\section{Theory and Methods}
\subsection{Closed-Shell Coupled Cluster Theory}
To avoid complicating the presentation that follows, we do not include singles in our algebraic formulation.
However, the inclusion of singles is straightforward.
In closed-shell coupled cluster with double excitations (CCD), we write the wave function as 
\begin{equation}
	\ket{\mathrm{CCD}} = e^{T_2}\ket{0},
\end{equation}
where  $\ket{0}$ is the restricted Hartree-Fock (RHF) reference, and the polynomial
$e^{T_2}$ is given by
\begin{equation}\label{eq:exp}
	e^{T_2} = 1 + T_2 + \frac{1}{2}T_2^2 + \frac{1}{6}T_2^3 + ...
\end{equation}
The cluster operator $T_2$ is spin adapted\cite{Scuseria1988} 
and creates double excitations:
\begin{equation}\label{eq:T2}
	T_{2} = \frac{1}{2}t_{ij}^{ab}E_a^iE_b^j,
\end{equation}
where $t_{ij}^{ab}$  refers to $t_{i_\uparrow j_\downarrow}^{a_\uparrow b_\downarrow}$ in a spin-orbital
formulation\cite{Scuseria1987,Scuseria1988}, and where
\begin{equation}\label{eq:E}
	E_p^q = p^\dagger_{\uparrow}q_{\uparrow} + p^\dagger_{\downarrow} q_{\downarrow}.
\end{equation}
Here, orbitals \textit{i j }(\textit{a b}) are occupied (unoccupied) in the RHF reference,
and summation over repeated indices is implied. 
We then construct a non-Hermitian, similarity-transformed Hamiltonian
\begin{equation}\label{eq:hbarcc}
	\overline{H}  =  e^{-T_2}He^{T_2},
\end{equation}
and obtain $T_2$ such that $\ket{0}$ is the right-hand eigenvector of $\overline{H}$
in the space spanned by $\ket{0}$ and the excitation manifold. Since $\overline{H}$ is non-Hermitian,
$\bra{0}$ is not its left-hand eigenvector, but we can
expand the left-hand
eigenvector $\bra{L}$ in the space spanned by $\bra{0}$ and
the excitation manifold as
\begin{subequations}\label{eq:Z}
\begin{align}
	\bra{L} = \bra{0}(1+Z_2), \\
	Z_{2} = \frac{1}{2}z_{ab}^{ij}E_i^aE_j^b.
\end{align}
\end{subequations}
The coupled cluster energy is the biorthogonal expectiation 
value of $\overline{H}$, and the amplitude equations needed 
to make $\ket{0}$ its right-hand eigenvector and $\bra{0}(1+Z)$ its
left-hand eigenvector are obtained by making the energy stationary:
\begin{subequations}\label{eq:CCeqs}
	\begin{align}
	E &=\braket{0|(1+Z_2)\overline{H}|0}, \\
		0 &= \frac{\partial E}{\partial t_{ij}^{ab}} = \frac{\partial E}{\partial z_{ab}^{ij}}.
	\end{align}
\end{subequations}

At convergence, the $z_{ab}^{ij}$ amplitudes do not contribute to the energy, and Eqs. \ref{eq:CCeqs}
reduce to the familiar CCD energy and amplitude equations
\begin{subequations}\label{eq:amps}
\begin{align}
	\braket{0|\overline{H}|0} &= E,\\
	\braket{_{ij}^{ab}|\overline{H}|0} &= 0, 
\end{align}
\end{subequations}
where $\ket{_{ij}^{ab}}$ is 
notation for referring to the space of 
doubly-excited determinants.\cite{Scuseria1988}

\begin{figure}
  \centering
  \includegraphics[width=1.0\textwidth]{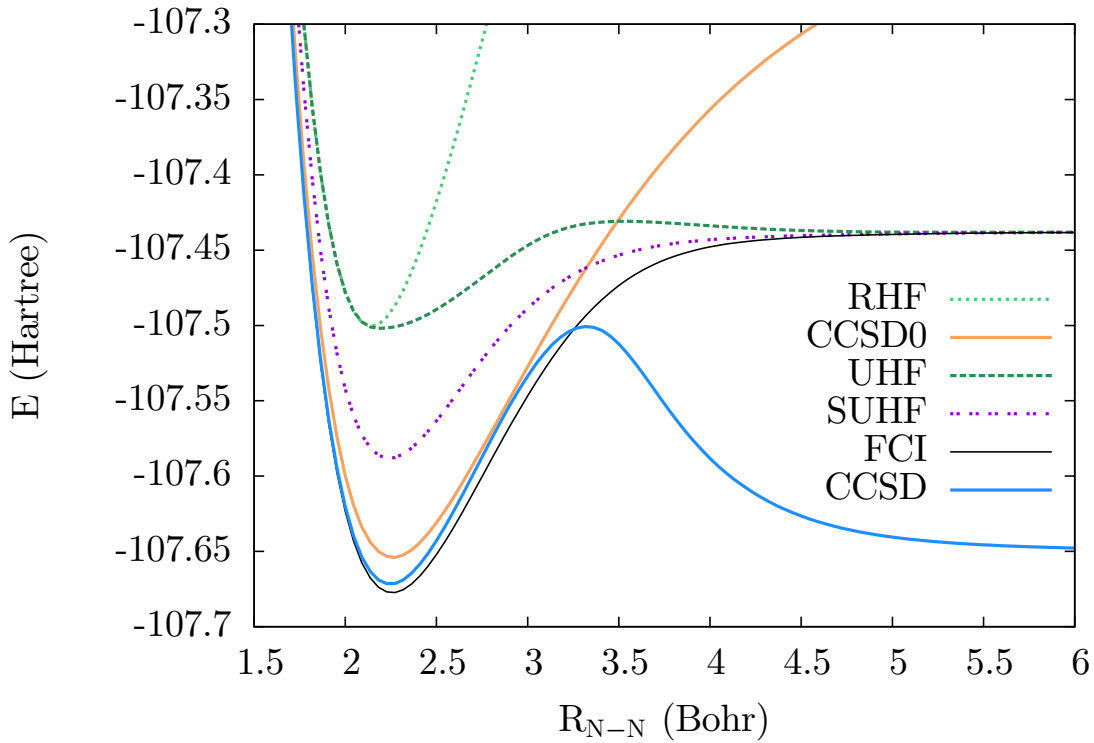}
  \caption{N$_2$ dissociation in STO-3G. CCSD is accurate near equilibrium, but overcorrelates at dissociation. SUHF
	   captures strong correlation at dissociation, but misses weak correlation near equilibrium. UHF
	   and SUHF dissociate to the correct limit due to the minimal basis. CCSD0 is protected from
	   breakdown, but sacrifices weak correlation throughout the curve.}
  \label{Fig:ccdbreakN2}
\end{figure}

With the addition of single excitations, CCSD accurately describes weakly-correlated systems. 
However, truncated coupled cluster methods on restricted Hartree-Fock references
are known to break down when systems take on multireference character, rendering
the mean field qualitatively incorrect. This breakdown is evidenced by the familiar
unphysical hump in the dissociation of N$_2$ in the STO-3G basis,
shown in Fig. \ref{Fig:ccdbreakN2}. Although CCSD is 
accurate near equilibrium, the method severely overcorrelates at stretched bond lengths. For comparison, 
we also include results from singlet-paired CCSD (CCSD0) in Fig. \ref{Fig:ccdbreakN2}. 
CCSD0 takes only the singlet-pairing channel of $T_2$ in Eq. \ref{eq:T2}, the effect of which is
to take the symmetric piece of the doubles amplitudes\cite{Gomez2016b}
\begin{equation}
	\sigma_{ij}^{ab} = \frac{1}{2}(t_{ij}^{ab} + t_{ij}^{ba}).
\end{equation}
This modification of $T_2$
protects the method from catastrophic failure at stretched bond lengths. 

In this work, we attempt to
repair CC by breaking away from the exponential ansatz. However, we would like to stress that it is
not the exponential ansatz that is inherently problematic, but its combination with solving the CC equations
projectively. Variational coupled cluster, in other words, does not break down due to strong correlation,
though it may undercorrelate severely. While variational coupled cluster\cite{Szalay1995,Kutzelnigg1998,
VanVoorhis2000B,Cooper2010,Evangelista2011}
is a fruitful
avenue of research, it is not a panacea, and introduces additional computational complexity.
We retain the standard practice of solving the CC equations projectively, and thus focus
our attentions on the form of the wavefunction ansatz.

\subsection{Projected Hartree Fock}
The SUHF wavefunction for closed shells is traditionally written as a singlet ($s=0$) spin projection operator $P$ acting on the $S^2$-broken UHF reference, $\ket{\phi}$. Thus,
\begin{equation}\label{eq:SUHFvar}
	\ket{\mathrm{SUHF}} =  P\ket{\phi},
\end{equation}
and the energy is obtained variationally\cite{Hoyos2012}
\begin{equation}\label{eq:SUHFE}
	E= \frac{\braket{\phi|P^\dagger H P|\phi}}{\braket{\phi|P^\dagger  P |\phi}} 
	= \frac{\braket{\phi| H P |\phi}}{\braket{\phi| P|\phi}},
\end{equation}
where we have used $ P^\dagger = P = P^2$ and $[H,P] = 0$.
Additionally, we employ a variation-after-projection approach, in which we deliberately break symmetries
and optimize the mean field in the presence of the projection operator.\cite{Scuseria2011,Hoyos2012} This procedure allows us to obtain the full SUHF dissociation curve in Fig. 
\ref{Fig:ccdbreakN2}, rather than obtaining improved energies only past the Coulson-Fischer point.

While we would like to try to combine SUHF and CC, it is difficult to see how to do so in a straightforward
manner because SUHF variationally solves for the energy as an expectation value, 
and CC solves for the energy and wavefunction 
projectively. Besides our own 
efforts, \cite{Qiu2016,Jacob2017,Qiu2017} there have also been other attempts to combine SUHF with residual 
correlation methods.\cite{Tsuchimochi2014,Tsuchimochi2015,Tsuchimochi2016}
In this work, we take a new approach, exploiting our recent formulation of SUHF for singlet states as a 
polynomial similarity transformation of particle-hole excitations out of a symmetry-adapted reference
determinant.\cite{Qiu2016,Qiu2017}
Although some of the details of the PoST SUHF formulation are not key to this work, we need to establish that CC and SUHF
can be written in a common language in order to justify modifying the CC equations based on the structure of the PoST SUHF
equations.
Briefly, we have shown that we can write the SUHF wavefunction as\cite{Qiu2016}
\begin{equation}\label{Eq:SUHF}
	\ket{\mathrm{SUHF}} =  e^{T_1}F(K_2)\ket{0}.
\end{equation}
In Eq. \ref{Eq:SUHF}, $T_1$ is the spin symmetry adapted single-excitation operator of standard coupled cluster
\begin{subequations}\label{Eq:E}
\begin{align}
	T_1 &= t_i^aE_a^i.
\end{align}
\end{subequations}
The polynomial $F(K_2)$ is given by\cite{Qiu2016}
\begin{subequations}\label{eq:F}
	\begin{align}
	F(K_2) &= 1 + K_2 + \frac{3}{10} K_2^2 + \frac{3}{70}K_2^3 + ...,
	\end{align}
\end{subequations}
which can be written as 
\begin{equation}\label{eq:FK2}
		F(K_2) = \frac{\mathrm{sinh}(\sqrt{6 K_2})}{\sqrt{6 K_2}}.
\end{equation}
The double excitation operator $K_2$ is 
\begin{subequations}\label{eq:K2}
\begin{align}
	K_2 &= \frac{1}{2}k_{ij}^{ab}E_a^i E_b^j, \\
	k_{ij}^{ab} &=-\frac{1}{3}(u_i^a u_j^b + 2 u_i^b u_j^a),
\end{align}
\end{subequations}
and $u_i^a$ are the adjustable parameters relating $\ket{0}$ to $\ket{\phi}$. More details
can be found in Refs. \cite{Qiu2016,Qiu2017}.

If we solve for the energy variationally, i.e.
\begin{equation}\label{eq:SUHFCCvar}
	E= \frac{\braket{0|F(K_2^\dagger)e^{T_1^\dagger} H e^{T_1} F(K_2) |0}}{\braket{0| 
	F(K_2^\dagger)e^{T_1^\dagger} e^{T_1} F(K_2) |0}},
\end{equation}
\begin{equation}\label{eq:SUHFCCvarstat}
	0  = \frac{\partial E}{\partial t_{i}^{a}} = \frac{\partial E}{\partial u_{i}^{a}},
\end{equation}
we get the variational SUHF energy,\cite{Qiu2016} so we know that the polynomial form of the SUHF wavefunction is exact. 
However, we solve the system `projectively', in a manner
similar to traditional coupled cluster.\cite{Qiu2016}

We thus write a similarity transformed Hamiltonian as (cf. Eq. \ref{eq:hbarcc})
\begin{equation}\label{eq:HPHF}
	\overline{H}_{\mathrm{PHF}} =  F^{-1}(K_2)e^{-T_1}He^{T_1}F(K_2),
\end{equation}
where the equations specifying the projective SUHF energy are
\begin{subequations}\label{eq:SUHFeqs}
\begin{align}
	E    &=\braket{0|(1+Z_1)G(L_2)\overline{H}_{\mathrm{PHF}}|0}, \\
	Z_1  &= \frac{1}{2}z_{a}^{i}E_i^a, \\
	L_2  &=-\frac{1}{6}(v_a^i v_b^j + 2 v_b^i v_a^j)E_i^a E_j^b,
\end{align}
\end{subequations}
where the $E_i^a$ excite when acting to the left,
and the stationarity conditions on the energy become
\begin{equation}\label{eq:SUHFstat}
	0  = \frac{\partial E}{\partial t_{i}^{a}} = \frac{\partial E}{\partial u_{i}^{a}} 
	   = \frac{\partial E}{\partial z_{a}^{i}} = \frac{\partial E}{\partial v_{a}^{i}}.
\end{equation}
In Eq. \ref{eq:SUHFeqs}, we define $G(L_2)$ such that $\bra{0}G(L_2)F^{-1}(K_2) \approx \bra{0}F^{\dagger}(K_2)$. 
To reproduce the SUHF variational energy, we need excitations in the bra up to the number of 
strongly correlated electrons.\cite{Qiu2016} Aside from needing a more complicated left-hand eigenvector for PoST SUHF 
to reproduce the variational SUHF energy, the basic mathematical structure of PoST SUHF described above is identical
to that of coupled cluster. In the next section, we explain how we combine PoST SUHF with CC.

\subsection{Attenuated Coupled Cluster}
\subsubsection{Similarity-Transformed Ansatz}
We posit that coupled cluster theory breaks down in the presence of strong correlation in part because it is trying
to mimic SUHF, which it cannot do, given the different polynomial forms of the two methods. When CC fails due to strong correlation,
it overcorrelates, while PoST SUHF does not. One explanation is that CC has a higher coefficient than PoST SUHF on terms quadratic and
higher in the amplitude equations, c.f. the $\frac{1}{2}$ on quadratic terms in Eq. \ref{eq:exp} vs 
the $\frac{3}{10}$ on quadratic terms in 
Eq. \ref{eq:F}. Larger coefficients results in more correlation, leading truncated CC to overcorrelate, partciularly
in the strongly correlated regime where some of the doubles amplitudes factorize and become large.\cite{Qiu2016}

Our general goal in this work is to write a wavefunction that incorporates information
from both the exponential similarity transformation of CC and the $sinh$ similarity transformation of SUHF. 
Our approach here is 
to write a new, double similarity transformation
\begin{equation}\label{eq:attHbar}
	\overline{H}_{\mathrm{attCC}} =  F^{-1}(K_2)e^{-S_2}He^{S_2}F(K_2),
\end{equation}
where the cluster operators have the forms
\begin{subequations}\label{eq:cluster}
\begin{align}
	S_2 = \frac{1}{2}s_{ij}^{ab}E_a^i E_b^j,\\
	K_2 = \frac{1}{2}k_{ij}^{ab}E_a^i E_b^j.
\end{align}
\end{subequations}
We would then solve for the energy and amplitudes projectively, as in coupled cluster theory:
\begin{subequations}\label{eq:attccen}
\begin{align}
	\braket{_{ij}^{ab}|\overline{H}_\mathrm{attCC}|0} &= 0,\\
	\braket{0|\overline{H}_{\mathrm{attCC}}|0} &=E.
\end{align}
\end{subequations}
The idea is that $S_2$ captures dynamical correlation via the coupled cluster exponential and
$K_2$ describes the spin collective mode and captures strong correlation via the SUHF $sinh$. We note that here 
we do not use
the more complicated left-hand eigenvector needed for PoST SUHF using $G(L_2)$, a potential shortcoming we discuss below.

Note that the total double-excitation operator is
\begin{equation}\label{eq:exctot}
	T_2 = S_2 + K_2,
\end{equation}
in terms of which the similarity transformed Hamiltonian can be written
\begin{subequations}\label{eq:attHbarbch}
\begin{align}
	\overline{H}_{\mathrm{attCC}} &=e^{-(T_2-K_2)}F^{-1}(K_2)H F(K_2)e^{T_2-K_2}  \\
	                              &=H + [H,T_2] + \frac{1}{2} [[H,T_2],T_2]  
	                                - \frac{1}{5}[[H,K_2],K_2] - \frac{2}{5}K_2[H,K_2] + ...
\end{align}
\end{subequations}
Thus, in practice, the energy of Eq. \ref{eq:attccen} uses the standard CCD expression, while the amplitude equations
are supplemented by two terms:
\begin{equation}\label{eq:attccenexp}
	\braket{_{ij}^{ab}|\overline{H}_\mathrm{attCC}|0} = \braket{_{ij}^{ab}|\overline{H}_\mathrm{CCD}|0}
	- \frac{1}{5}\braket{_{ij}^{ab}|[[H,K_2],K_2]|0} - \frac{2}{5}\braket{_{ij}^{ab}|K_2[H,K_2]|0}.
\end{equation}
Everything up to this point is rigorous. What is approximate is how we obtain $K_2$ from $T_2$, determining
the spin collective mode on the fly. We describe this procedure next.

\subsubsection{Determining the Spin Collective Modes}
In order to identify the
$s_{ij}^{ab}$ and
$k_{ij}^{ab}$ amplitudes, we look in the CCD $t_{ij}^{ab}$ amplitudes because
spin projection is a collective phenomenon, yielding large amplitudes to all orders
that should not be neglected in the strongly correlated regime.\cite{Degroote2016,Qiu2016} 
We write the CCD amplitudes as
\begin{equation}\label{eq:T2part}
	t_{ij}^{ab} = k_{ij}^{ab} + s_{ij}^{ab},
\end{equation}
and identify the spin collective mode of the CCD amplitudes by noting 
that the object $k_{ij}^{ab} - 2k_{ij}^{ba}$ constructed with SUHF $k_{ij}^{ab}$ amplitudes from
Eq. \ref{eq:K2} is described
by a single eigenvector, or collective mode, and factorizes exactly into single excitations, i.e.
\begin{equation}\label{eq:K2fact}
	k_{ij}^{ab} - 2k_{ij}^{ba} = u_{i}^{a}u_{j}^{b}.
\end{equation}
An analogus object, constructed with $t_{ij}^{ab}$ amplitudes from CC, rather than $k_{ij}^{ab}$ amplitudes
from SUHF, does not factorize exactly. 
However, we can 
construct a matrix $U_2$ whose elements are
\begin{equation}\label{eq:U}
	U_{ai,bj} = t_{ij}^{ab} - 2 t_{ij}^{ba}.
\end{equation}
We then diagonalize $U_2$ along particle-hole lines
\begin{equation}\label{eq:Udiag}
	U_{ai,bj} = \sum\limits_n\lambda_n V_{ai,n}V_{bj,n},
\end{equation}
where $\lambda$ are the eigenvalues and $V$ are the eigenvectors.

\begin{figure}
  \centering
  \includegraphics[width=1.0\textwidth]{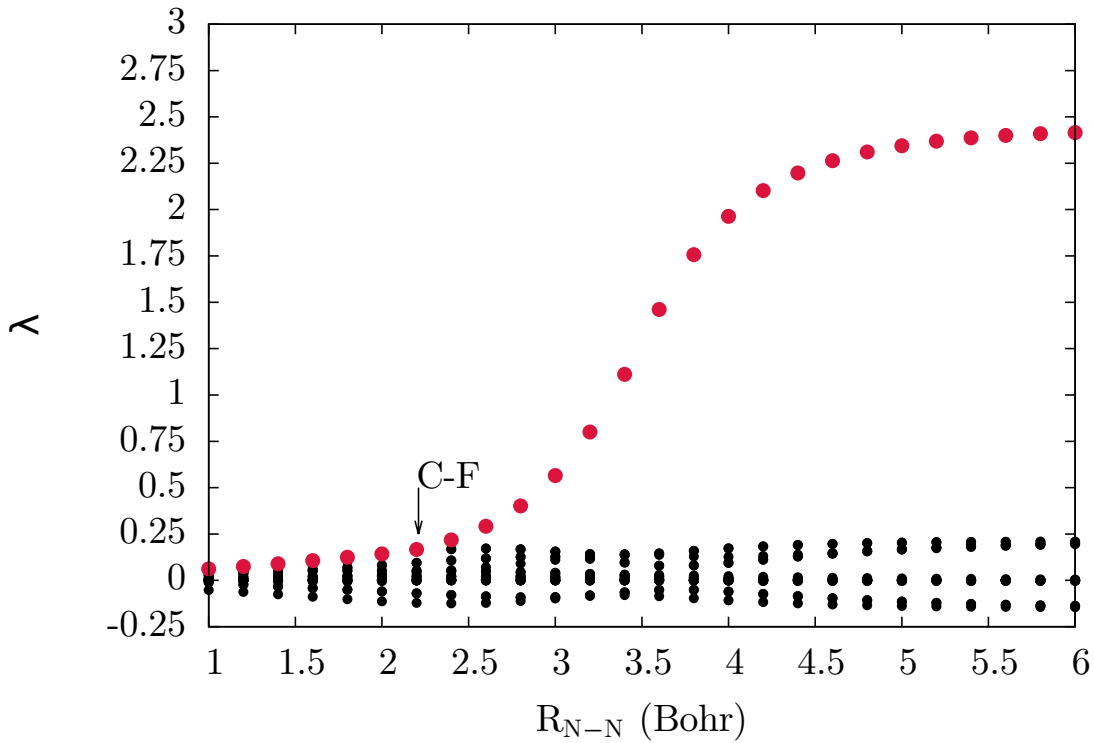}
  \caption{N$_2$ STO-3G eigenvalues of $U_2$ built from CCD amplitudes. Past the Coulson-Fischer point,
  a single large eigenvalue spin collective mode (shown in red) dominates.}
  \label{fig:N2Eig}
\end{figure}

The eigenvalues of $U_{ai,bj}$ from Eq. \ref{eq:U}, constructed from 
traditional CCD $t_{ij}^{ab}$ amplitudes for the STO-3G 
dissociation of N$_2$, are shown in Fig. \ref{fig:N2Eig}.
Although the Hartree-Fock reference is constrained to be restricted
throughout the range of bond lengths, we have labeled the bond length at which 
spontaneous RHF spin symmetry breaking occurs (the Coulson-Fischer point)
as `C-F'. For bond lengths below C-F, the eigenvalues of $U_2$ are all small, and
CC accurately describes the weakly-correlated system. However, once we pass the critical point, and the 
system becomes strongly correlated,
a single collective mode begins to dominate the ph-ph spectrum of $T_2$. In other words, the CC wavefunction
seems to be trying to mimic the structure of the SUHF amplitudes in Eq. \ref{eq:K2fact}, which
require a different polynomial.
Treating this collective mode
with the exponential of CC is partly responsible for the breakdown of the method. 
Eliminating this mode seems to protect CC from breakdown, but severely undercorrelates, motivating our
attenuation, rather than elimination, of the collective mode.
We propose to treat the amplitudes
corresponding to this spin collective mode using the SUHF $sinh$ polynomial.

Following our discussion of Fig. \ref{fig:N2Eig}, and comparing Eqs. \ref{eq:K2fact} and \ref {eq:Udiag}, we identify the largest eigenvalue of $U_2$ as the spin
collective mode and write the collective and non-collective parts of $U_2$
\begin{subequations}\label{eq:Ucoll}
\begin{align}
	U_{ai,bj}^c &= \lambda_{\max} V_{ai,n_{\lambda_{\mathrm{max}}}}V_{bj,n_{\lambda_{\mathrm{max}}}},\\
	U_{ai,bj}^{nc} &=  U_{ai,bj} -  U_{ai,bj}^c 
\end{align}
\end{subequations}
where $U^c$ and $U^{nc}$ refer to the collective- and noncollective blocks, respectively.
Using Eqs. \ref{eq:Ucoll}, we reconstruct $k_{ij}^{ab}$ amplitudes, i.e. CC amplitudes that want to mimic SUHF,
and $s_{ij}^{ab}$ amplitudes, which are the rest of the CC amplitudes that simply
capture weak correlation (compare to Eq. \ref{eq:K2})
\begin{subequations}\label{eq:ksamps}
\begin{align}
	k_{ij}^{ab} &= -\frac{1}{3}(U_{ai,bj}^c + 2U_{aj,bi}^c), \\
	s_{ij}^{ab} &= -\frac{1}{3}(U_{ai,bj}^{nc} + 2U_{aj,bi}^{nc}) = t_{ij}^{ab} - k_{ij}^{ab}.
\end{align}
\end{subequations}

\subsection{Computational Details}
All calculations on the Hubbard and pairing Hamiltonians were performed using in-house code. The
calculations on the Hubbard Hamiltonian all use periodic boundary condtions (PBC). Coupled
cluster calculations on the Hubbard model use the RHF plane wave basis.
Molecular Hartree-Fock and standard coupled cluster calculations were done in \textit{Gaussian 09},\cite{g09e1} while the attCC calculations were performed
using in-house code. Standard extrapolation techniques are used
to accelerate convergence.\cite{Scuseria1986b}
Our calculations are done without point group symmetry. 
Thus, though we have $S=0$ SCF wavefunctions, they may break other symmetries, e.g. point group.
Molecular bond lengths
are in units of Bohr.
We use cartesian $d$ functions and work in relatively small basis sets\cite{Feller1996,esml}
in order to exacerbate the effects of strong correlation and
emphasize the deficiencies of 
standard coupled cluster in this regime. 

\section{Results}
\subsection{Molecules}
\begin{figure}
  \centering
  \includegraphics[width=1.0\textwidth]{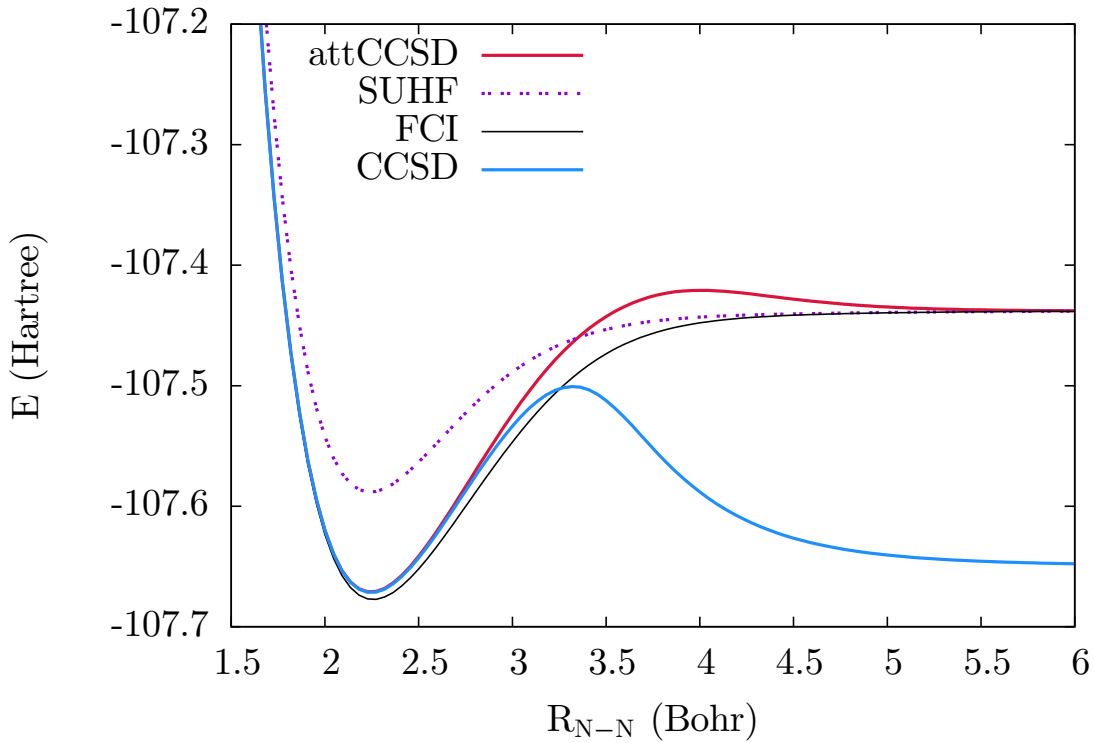}
  \caption{N$_2$ dissociation in STO-3G. CCSD overcorrelates dramatically at dissociation. SUHF dissociates correctly, but misses 
  correlation at equilibrium. attCCSD captures correlation at both equilibrum and dissociation.}
  \label{fig:N2sto3g}
\end{figure}

We first test attCC on some small molecular systems.
Results for the dissociation of N$_2$ in the STO-3G basis are shown in Fig. \ref{fig:N2sto3g}, where
we plot total energies as function of bond length. As discussed earlier, the breakdown of CCSD is quite pronounced, as CCSD
turns over near 3.2 bohr and overcorrelates dramatically at dissociation. 
As a result of the small basis, SUHF is exact at dissociation, but near
equilibrium we can see that SUHF misses much of the dynamical correlation. In a sense, attCCSD offers the best of
both worlds, giving energies comparable to CCSD near equilbrium and dissociating to the SUHF limit. 
Although there is a small bump in the attCCSD dissociation curve, it is qualitatively, and nearly quantitatively,
correct.

\begin{figure}
  \centering
  \includegraphics[width=1.0\textwidth]{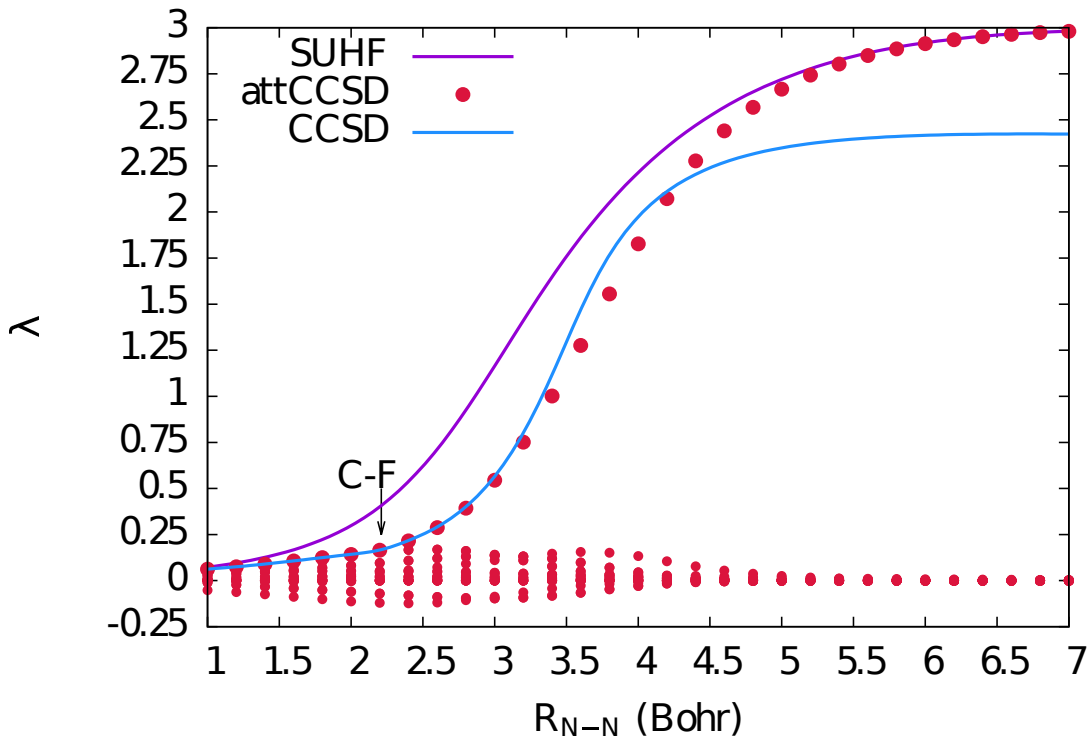}
  \caption{N$_2$ STO-3G eigenvalues of $U_2$. Attenuation gives the SUHF spectrum at dissociation.}
  \label{fig:N2attCCDEig}
\end{figure}

In Fig. \ref{fig:N2attCCDEig}, we plot the eigenvalues of $U_2$ (see
Eq. \ref{eq:Ucoll}), complementing Fig. \ref{fig:N2Eig}. Here, $U_2$ is constructed with standard CSCD
and attCCSD doubles amplitudes to show what happens to the spectrum after attenuation. For clarity,
we have only shown the largest eigenvalue from CCSD. In the PoST formulation
of SUHF, the eigenvalues are absorbed into the definition of $u$ in Eq. \ref{eq:K2fact}, so
for SUHF in Fig. \ref{fig:N2attCCDEig}, we plot $u^\dagger u = \lambda$. The largest eigenvalue of attCC goes
to the SUHF limit, and is unaffected by the addition of singles. The non-collective modes of attCC
go to zero. Compared with Fig. \ref{fig:N2Eig}, we see that the effect of attenuation in the strongly-correlated limit
reproduces the spectrum of the SUHF wavefunction: the collective mode goes to SUHF, and
the non-collective modes go to zero. This result also explains why attCC in general undercorrelates:
when systems are too strongly-correlated, attCC looks too much like SUHF, and therefore lacks dynamic correlation.
It is probable that residual dynamic correlation here originates from connected triples.

\begin{figure}
  \centering
  \includegraphics[width=1.0\textwidth]{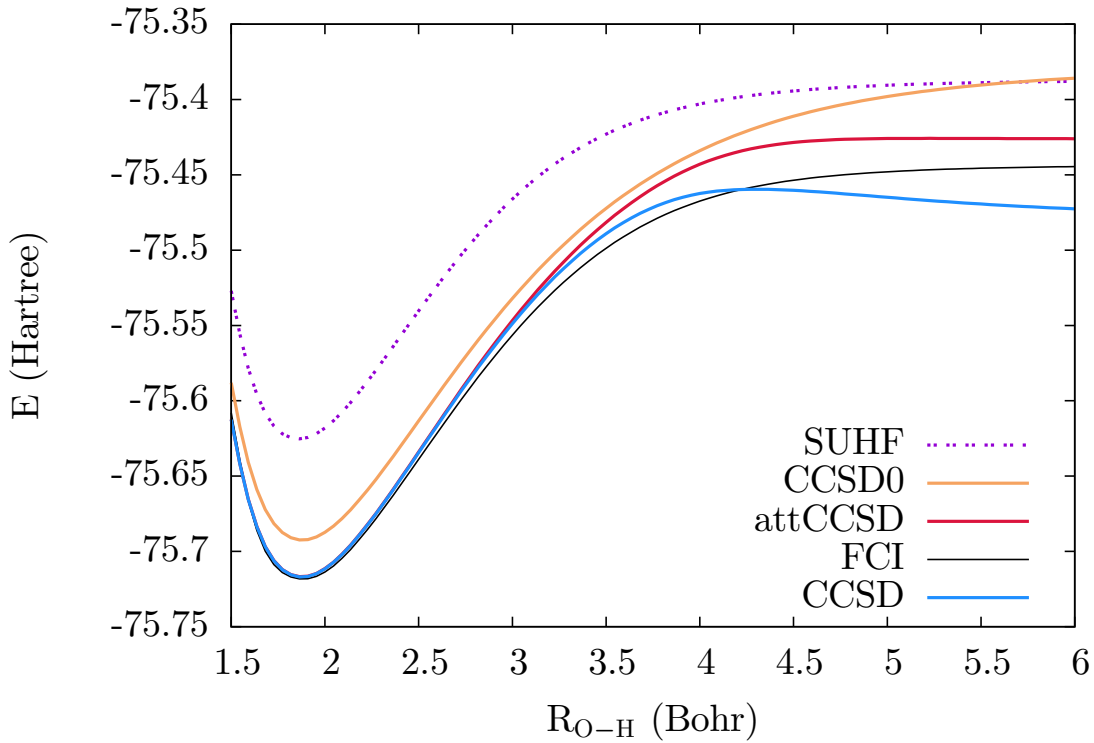}
  \caption{H$_2$O dissociation in 3-21G, $\theta_{\mathrm{H-O-H}} = 104.52\degree$. CCSD overcorrelates
  at dissociation. CCSD0 and SUHF are well behaved but sacrifice dynamical correlation. attCCSD captures
   nearly all dynamical correlation at equilibrium and improves on SUHF and CCSD at dissociation.}
  \label{fig:H2O}
\end{figure}

Results for the symmetric dissociation of water with an H-O-H angle of $104.52\degree$\cite{Bytautas2015}
in
the 3-21G basis are shown in Fig. \ref{fig:H2O}, where we plot
total energies as a function of the O-H bond length. We include single excitations, although
results without singles are comparable in the minimal basis calculations shown here. At equilibrium, CCSD
is very close to full configuration interaction (FCI), the exact result in this basis set. However, as 
the bond stretches, CCSD
breaks down, turning over at around 4 bohr and overcorrelating at dissociation. CCSD0 is protected from
breakdown, but sacrifices a large amount of dynamical correlation across all bond lengths. SUHF is well behaved
throughout the curve, but also sacrifices dynamical correlation, particularly at equilibrium. 
At quilibrium, attCCSD is
nearly identical with CCSD, giving energies better than SUHF in this
regime. At dissociation, attCCSD is protected from the breakdown suffered by CCSD and gives energies
superior to both CCSD0 and SUHF.

\begin{figure}
  \centering
  \includegraphics[width=1.0\textwidth]{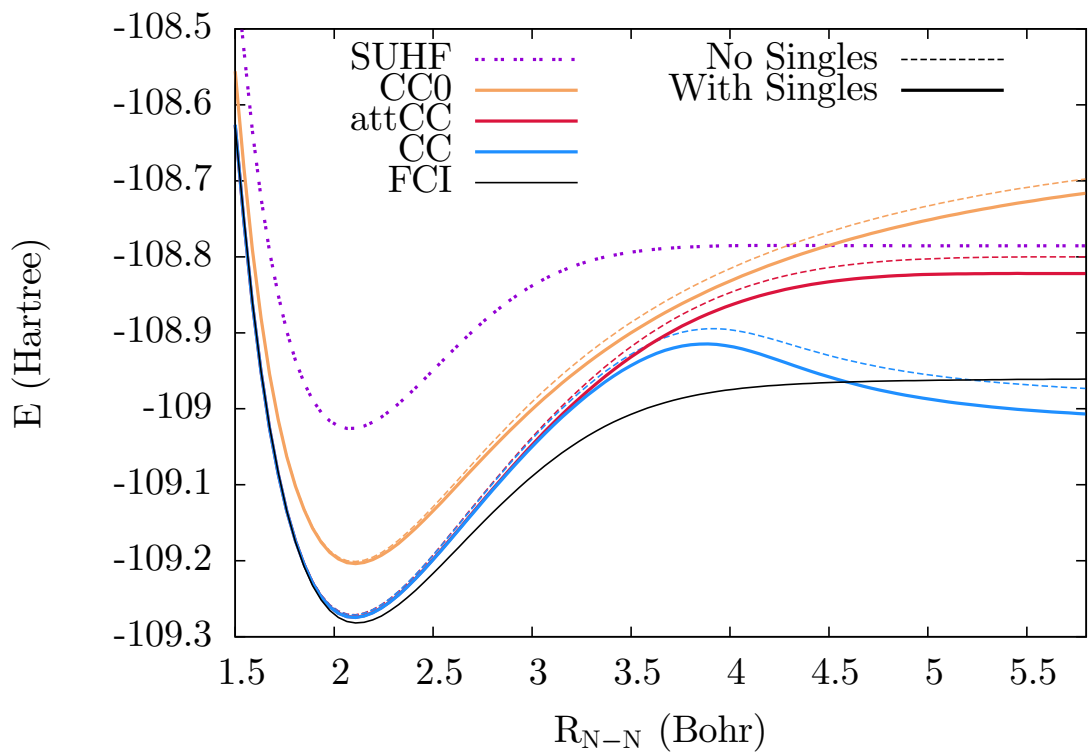}
  \caption{N$_2$ dissociation in cc-pVDZ. CC breaks down, overcorrelating at dissociation. CC0 is protected
  from breakdown, but misses weak correlation. SUHF has the correct shape, but misses weak correlation. attCCSD
  is accurate at equilibrium and improves on SUHF at dissociation, but still misses correlation
  from the intermediate region outward.}
  \label{fig:N2dz}
\end{figure}

Lastly, we look at the dissociation of N$_2$ in the larger cc-pVDZ basis\cite{Dunning1989} in
Fig. \ref{fig:N2dz}. Calculations including singles and
doubles are shown with solid lines; the corresponding doubles-only calculations are shown with dotted lines. 
The FCI results are all-electron, but use spherical $d$-functions, which raises the
energy slightly.
As in the minimal basis case, standard CC breaks down, turning over around 3.6 bohr and overcorrelating
at dissociation. CC0 once again is protected from breakdown, but sacrifices dynamical correlation. SUHF gives
the correct shape of the dissociation curve. However, there is more dynamical correlation in this larger 
basis. SUHF fails to capture this weak correlation, and gives poor energies overall. Once again, attCC gives
excellent energies at equilibrium, is protected from breakdown at dissociation 
and improves on SUHF energies throughout the curve.
However, even with the addition of singles,
attCCSD misses a fair amount of correlation at dissociation, a failure we address in more detail in the
Discussion section.

\subsection{Hubbard}
We now look at the 1-dimensional Hubbard model\cite{Hubbard1963} at
half filling with periodic boundary conditions (PBC).  
The Hamiltonian is given by
\begin{equation}\label{Hubbard}
	H = -t \sum \limits_{j,\sigma}(c^\dagger_{j+1,\sigma}c_{j,\sigma}+c^\dagger_{j,\sigma}c_{j+1,\sigma})
	+ U \sum\limits_{j}n_{j\uparrow}n_{j\downarrow},
\end{equation}
where $c^\dagger_{j,\sigma}$ and $c_{j,\sigma}$ create and annihilate an electron with spin $\sigma$ on
site $j$, respectively, and $n_{j\sigma} = c^\dagger_{j,\sigma} c_{j,\sigma}$, the standard number operator for electrons of spin $\sigma$ on site $j$. The
parameter $t$ allows electrons to hop between adjacent sites. The parameter
$U > 0 $ represents repulsion of electrons on the same site. The model is well studied and loosely corresponds
to a minimal basis chain of hydrogen atoms where the ratio $U/t$ is analogous to the interatomic bond distance.
For large values of $U/t$, the system is strongly correlated. An attractive feature of the 1D Hubbard model 
with PBC is 
that exact results are available via a Bethe ansatz solution.\cite{Bethe1931,LiebWu}
RHF for this system yields plane waves,
as opposed to the dimerized basis which is spin symmetry adapted, but translational symmetry broken.
Due to momentum symmetry, single excitations are zero.

\begin{figure}
  \centering
  \includegraphics[width=1.0\textwidth]{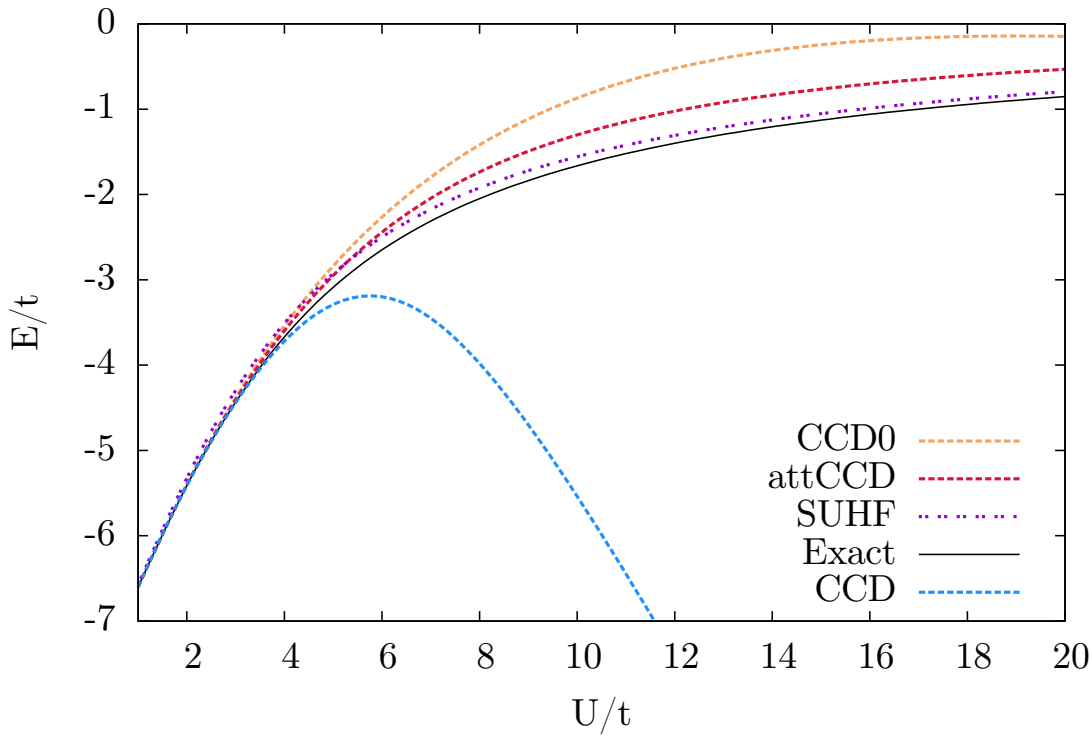}
  \caption{6-site Hubbard in the plane-wave basis. Singles are zero by momentum symmetry. CCD breaks down, turning over around $U/t=6$. SUHF is good everywhere. attCCD
           improves over both CCD and CCD0.}
  \label{fig:6site}
\end{figure}

Results for the 6-site Hubbard model are shown in Fig. \ref{fig:6site}. For small values
of $U/t$, CCD is essentially exact. However, CCD begins to overcorrelate near $U/t = 4$, eventually
turning over and heading toward negative infinity. Singlet-paired coupled cluster (CCD0), which
we have found to be quite robust in molecular systems,\cite{Bulik2015,Gomez2016b}
is nearly exact for small values of $U/t$, but begins to plateau around $U/t=20$, eventually turning over. 
SUHF is well-behaved everywhere, but misses correlation throughout the curve, particularly for intermediate
values of $U/t$. 
For small values of $U/t$,
attCCD is nearly equivalent to CCD, and therefore
superior to SUHF for $U/t \lesssim 5$. As a correction to CCD, attCCD at large 
values of $U/t$ is remarkable, as the relatively small modification we have made 
protects the method from breakdown. However, SUHF energies are much better than attCCD energies as $U/t$ becomes
large.

\begin{figure}
  \centering
  \includegraphics[width=1.0\textwidth]{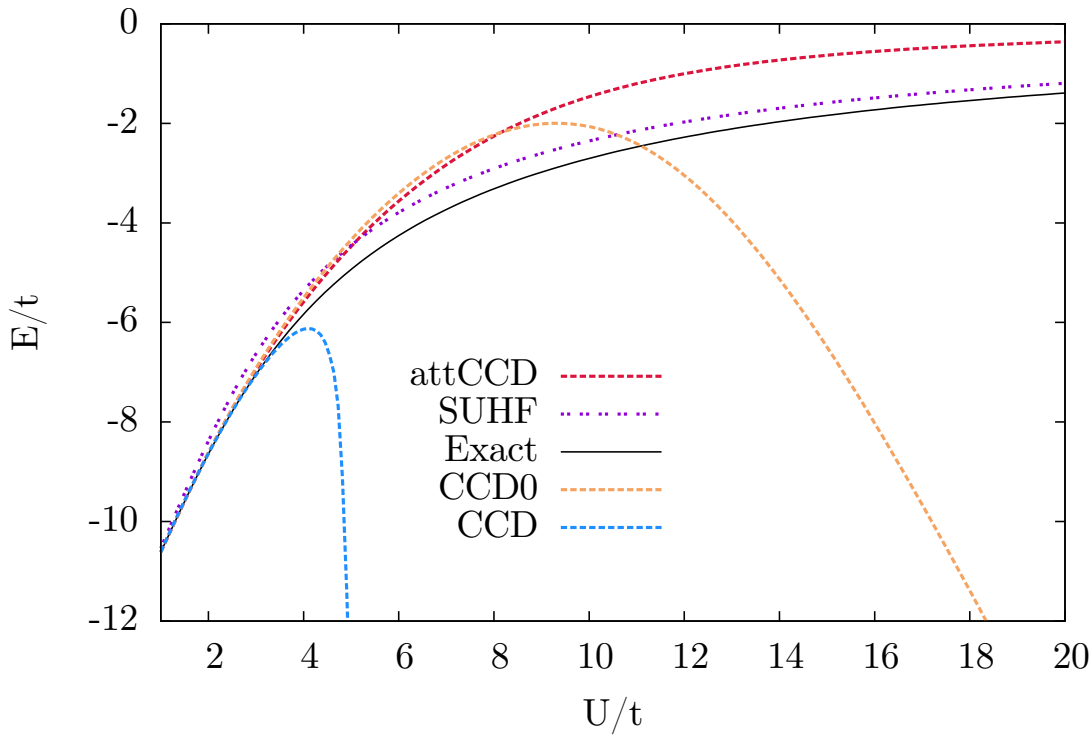}
  \caption{10-site Hubbard. CCD breaks down, turning over sharply around $U/t=4$. CCD0 turns over
  around $U/t=8$. SUHF is good everywhere. attCCD improves dramatically over both CCD and CCD0.}
  \label{fig:10site}
\end{figure}

\begin{figure}
  \centering
  \includegraphics[width=1.0\textwidth]{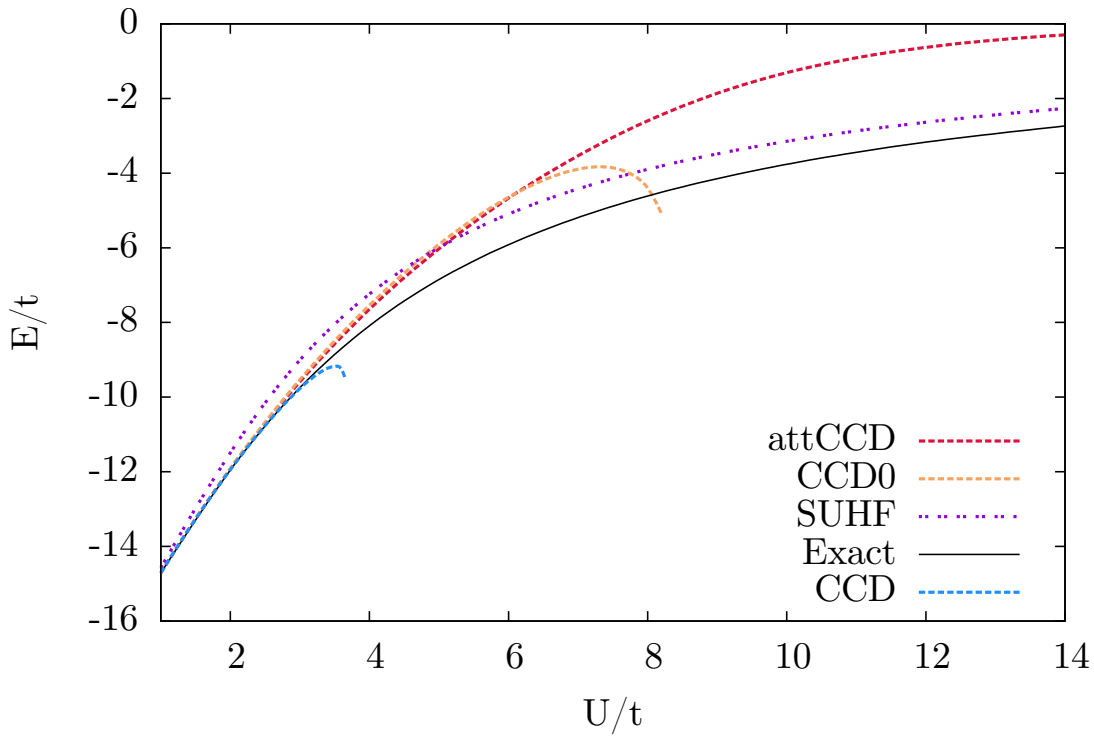}
  \caption{14-site Hubbard. CCD breaks down, turning over sharply and becoming
	  difficult to converge before $U/t=4$. CCD0 turns over
  around $U/t=7$ and stops converging at $U/t=8$. SUHF is reasonable everywhere. attCCD improves dramatically over both CCD and CCD0.}
  \label{fig:14site}
\end{figure}

Figs. \ref{fig:10site} and \ref{fig:14site} show results
for 10- and 14-site Hubbard rings, respectively. The same qualitative features in our 6-site results
are observed here as well. For small values of $U/t$, attCCD is quite accurate.
As the system size increases, attCCD becomes less accurate at large values of $U/t$, relative to 
SUHF. However, as the system size increases, traditional CCD becomes progressively worse,
turning over and failing to converge for smaller $U/t$. Although attCCD becomes less accurate
as system size increases, it is also protected from the increasingly severe breakdown
of CCD. We should note that
SUHF is not size extensive and reverts to the same energy as UHF for large systems, an effect
we seem to be seeing in the attCCD results.

\subsection{Pairing Attenuation}
We have focused primarily on incorporating $S^2$ projection into CC in this work. 
However, we can combine CC with the projection
of yet other symmetries, such as particle number. Particle number may be spontaneously broken
in attractive Hamiltonians in analogy with the breaking of spin in repulsive Hamiltonians, 
but is more straight forward to attenuate
than other repulsive Hamiltonian symmetries, e.g. $S_z$.

Consider the pairing or reduced BCS Hamiltonian, which
is a simple model of superconductivity and is a good system to test our methods because
it is very challenging for traditional coupled cluster:\cite{Henderson2014b,Degroote2016}
\begin{subequations}\label{eq:pairing}
\begin{align}
	H &= \sum\limits_{p} \epsilon_pN_p - G \sum\limits_{p,q}P^\dagger_pP_q,\\
	N_p &= c^\dagger_{p\uparrow}c_{p\uparrow} +c^\dagger_{p\downarrow}c_{p\downarrow} ,\\
	P^\dagger_p &= c^\dagger_{p\uparrow}c^\dagger_{p\downarrow},\\
	P_p &= c_{p\uparrow}c_{p\downarrow}.
\end{align}
\end{subequations}
Here, the sums are over single-particle levels of energy $\epsilon_p$ and the Hamiltonian has interaction strength $G$. As
with 1D Hubbard, exact results are available.\cite{Richardson1963,Richardson1966} As
$G$ increases, the HF reference tends toward a particle-number broken BCS solution at interactions
$G$ larger than the critical point $G_{\mathrm{c}}$. 

The number-projected wavefunction (PBCS)
can be written as a Bessel polynomial of pair excitations out of the number-preserving Hartree-Fock reference\cite{Degroote2016}
\begin{subequations}\label{eq:PHFB}
\begin{align}
	\ket{\mathrm{PBCS}} &= (1 + T_2 + \frac{1}{4} T_2^2 + \frac{1}{36} T_2^3 + ... )\ket{0},\\
	T_2 &= t_{ii}^{aa}P^\dagger_aP_i.
\end{align}
\end{subequations}
where the quadratic terms now carry a coefficient of $\frac{1}{4}$.
Now, the pairing amplitudes factorize directly along particle-particle/hole-hole (pp-hh) lines as
\begin{equation}\label{eq:pairfact}
	t_{ii}^{aa} = x_iy^a.
\end{equation}
Due to a symmetry of the Hamiltonian, $t_{ij}^{ab} = 0$ if $i\ne j$ or $a \ne b$ for this system. Single and triple
amplitudes are zero as well.
Following the above procedure, we thus build 
\begin{equation}
	U_{ii,aa} = t_{ii}^{aa},
\end{equation}
using the $t_{ij}^{ab}$ amplitudes from CCD. We then perform a singular
value decomposition of $U_{ii,aa}$ along pp-hh lines and identify the largest
singular value as the pairing collective mode. Going back to the CCD equations, we treat the terms 
quadratic in the pairing collective mode amplitudes with the coefficient of $\frac{1}{4}$ from
the Bessel PBCS polynomial.

\begin{figure}
  \centering
  \includegraphics[width=1.0\textwidth]{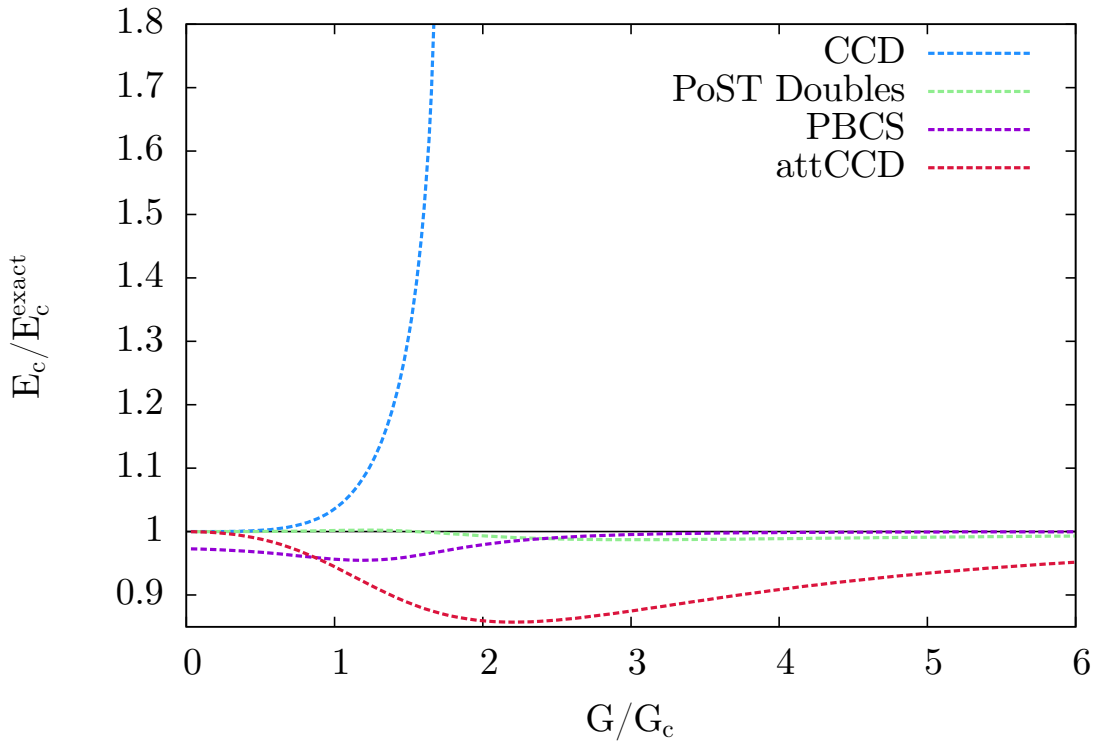}
  \caption{12-site Pairing Hamiltonian. CCD overcorrelates, eventually going complex. PBCS is well
  behaved everywhere. attCCD is protected from breakdown at large $G$. PoST Doubles gives excellent results, but has much higher computational scaling than attCCD.}
  \label{fig:12site}
\end{figure}

We show results for the 12-level pairing Hamiltonian
in Fig. \ref{fig:12site}. Here, we plot the fraction of correlation energy recovered as a function of $G/G_c$.
CCD is very accurate for small values of $G/G_c$, but overcorrelates drastically past the
symmetry breaking point, eventually going complex.\cite{Henderson2014b} Attenuated coupled cluster is comparable
to CCD for small $G/G_c$. Although attCCD misses some correlation, especially
in the intermediate region, it is protected from the breakdown of CCD. 

As a comparison, we also show results from PoST Doubles,\cite{Degroote2016} which is a polynomial similarity transformation of double
excitations that interpolates between CCD and PBCS.
To do so,
PoST Doubles
gives $c_2 = \frac{1}{2} $ (coupled cluster) for small $G$ in the pairing Hamiltonian
and $c_2 = \frac{1}{4}$ (PBCS) for large $G$, determining the value of $c_2$ by minimizing
the quadruples $R_4$ residual.\cite{Degroote2016} 
Because it must minimize the $R_4$ residual, PoST Doubles has much higher computational scaling than
attCCD and is difficult to apply to realistic systems.
While the results for attCCD are not as good as those for PoST Doubles, 
attenuation is simpler and requires less computational effort.

Attenuated coupled cluster is not an interpolation, but
applies the PBCS Bessel polynomial to one block of the amplitudes and the CC exponential polynomal
to the the rest of the amplitudes (i.e. we apply the attenuation only in a `factorized' eigenvector
basis). It might be possible to make
attCCD more in the spirit of PoST Doubles by introducing an interpolation scheme 
for the coefficient of the pairing collective mode,
rather than applying a blanket $c_2 = \frac{1}{4}$.

\section{Discussion}
In this work, we have demonstrated a simple method of combining the PoST singles formulation of 
spin symmetry projection
with the exponential form of coupled cluster that has shown some promising initial results. 
In contrast with our previous work on singlet-paired coupled 
cluster, here we attenuate, rather than sever, the problematic interaction between channels in the 
CC equations, and do so along particle-hole/particle-hole lines instead of particle-particle/hole-hole
lines as in Ref. \cite{Bulik2015}.

It is important to emphasize that directly replacing  the quadratic coefficient $c_2 = \frac{1}{2}$ 
with $c_2 = \frac{3}{10}$ and adding the unlinked terms
in a CCD code does not give SUHF. As discussed above, such a procedure instead gives the projective equations 
\begin{subequations}\label{eq:PSUHF}
\begin{align}
	\braket{0|\overline{H}_{\mathrm{PHF}}|0} &=E, \\
	\braket{_{ij}^{ab}|\overline{H}_{\mathrm{PHF}}|0} &= 0, 
\end{align}
\end{subequations}
which is manifestly not equal to the SUHF expectation value of Eq. \ref{eq:SUHFeqs},
because of the inadequate bra state.
We might refer to the method described in Eq. \ref{eq:PSUHF} as `projective' SUHF (pSUHF), and
it is equivalent to spin attenuating all modes of $t_{ij}^{ab}$, rather than just the collective mode
attenuated in attCC. We have tested pSUHF, and though it is protected from breakdown due to
strong correlation, like attCC, it gives worse energies. In a sense, the more the wavefunction
can look like coupled cluster, withought reintroducing the breakdown, the better. To this end,
attenuating only the spin collective mode seems superior to attenuating all modes. Indeed,
we might be able to improve upon attCC by introducing an interpolation scheme 
for the coefficient of the spin collective mode, as described above
for the pairing Hamiltonian. 
We are keenly interested in such a procedure,
but determining how to optimize $c_2$ here is not straightforward.

Another shortcoming of attCC that will require some thought
is the nature of the left-hand eigenvector. Currently, attCCD uses the 
projective coupled cluster-style amplitude and energy equations of Eq. \ref{eq:attccen}.
As PoST SUHF in Eq. \ref{eq:SUHFeqs} requires the left-hand eigenvector to produce the variational SUHF energy,
a composite method of CC and SUHF may also need a more complicated left-hand side. The nature
of such a formalism is the subject of ongoing work in our group, and we are optimistic about
the promise of simultaneously describing weak and strong correlation
with a combination of coupled cluster and symmery projection.

\section{Acknowledgements}
J.A.G. thanks Jacob Wahlen-Strothman for useful discussions, 
Dr. Matthias Degroote for help with the pairing Hamiltonian,
Ethan Qiu for some of the exact results and
Dr. Carlos Jim{\'e}nez-Hoyos for his help with SUHF.
This material is based on work supported by the National Science Foundation
(CHE-1462434).
G.E.S. is a Welch Foundation Chair (C-0036).
J.A.G. gratefully acknowledges the support of the National Science Foundation Graduate Research Fellowship Program (DGE-1450681). 

\bibliographystyle{tfo}
\bibliography{attcc}
\end{document}